\documentclass[%
twocolumn,
superscriptaddress,
 amsmath,amssymb,
 aps,
 pra,
]{revtex4-1}

\usepackage{placeins}
\usepackage{SIunits}
\usepackage{amsmath}    
\usepackage{graphicx}   
\usepackage{verbatim}   
\usepackage{color}      
\usepackage{subfigure}  
\usepackage{hyperref}   
\usepackage[normalem]{ulem} 
\usepackage{multirow}
\usepackage[normalem]{ ulem } 
\usepackage{soul}             
\usepackage{xspace}

\begin{document}

\title{Theoretical study of the I$^+$ + I$^-$ mutual neutralization reaction}

\author{Sylvain~Badin}
\affiliation{Université de Lille, CNRS, UMR 8523 - PhLAM - Physique des Lasers, Atomes et Molécules, F-59000 Lille,  France}
\affiliation{Sorbonne Universit\'{e}, CNRS, Laboratoire de Chimie Physique Mati\`{e}re et Rayonnement, UMR 7614, F-75005 Paris, France}
\author{Xiang~Yuan}
\affiliation{Université de Lille, CNRS, UMR 8523 - PhLAM - Physique des Lasers, Atomes et Molécules, F-59000 Lille,  France}
\affiliation{Department of Chemistry and Pharmaceutical Science, Faculty of Science, Vrije Universiteit Amsterdam, de Boelelaan 1083, 1081 HV Amsterdam, The Netherlands.}
\author{Pierre-Louis~Bourgeois}
\affiliation{Sorbonne Universit\'{e}, CNRS, Laboratoire de Chimie Physique Mati\`{e}re et Rayonnement, UMR 7614, F-75005 Paris, France}
\author{Andre~Severo~Pereira~Gomes}
\affiliation{Université de Lille, CNRS, UMR 8523 - PhLAM - Physique des Lasers, Atomes et Molécules, F-59000 Lille,  France}
\author{Nicolas~Sisourat}
\affiliation{Sorbonne Universit\'{e}, CNRS, Laboratoire de Chimie Physique Mati\`{e}re et Rayonnement, UMR 7614, F-75005 Paris, France}
\begin{abstract}

We have computed the cross sections of the mutual neutralization reaction
between I$^{+}$ and I$^{-}$ for a collision energy varying from
0.001 eV to 50 eV. These cross sections were obtained using the adiabatic potential energy curves of the I$_{2}$
system computed with a direct relativistic Multi-Reference Configuration Interaction method and a semi-classical approach (i.e. Landau Zener Surface Hopping). We report the cross sections towards the following neutral sates
: $\text{I}(^{2}P_{3/2})+\text{I}(^{2}P_{3/2})$, $\text{I}(^{2}P_{3/2})+\text{I}(^{2}P_{1/2})$, $\text{I}(^{2}P_{1/2})+\text{I}(^{2}P_{1/2})$
and $\text{I}(5p^{4}6s)+\text{I}(^{2}P_{3/2})$. We also discuss the cross sections towards
the two following excited ionic states : $\text{I}^{-}(^{1}S_{0})+\text{I}^{+}(^{3}P_{0})$
and $\text{I}^{-}(^{1}S_{0})+\text{I}^{+}(^{3}D_{2})$. The results of these calculations are
in accordance with recent experimental measurements conducted in the double ion
ring DESIREE in Stockholm. These results can be used to model iodine plasma
kinetics and thus to improve our understanding of the latter.

\end{abstract}

\maketitle

\section{Introduction}
\label{sec:intro}

The mutual neutralization (MN) of two oppositely charged ions is a central reaction taking place in electronegative plasmas. The latter are found in e.g. the lower ionosphere~\cite{shuman_ambient_2015}, flames~\cite{fialkov_investigations_1997}, interstellar medium~\cite{smith_mutual_1976,smith_ion_1992} and in excimer lasers~\cite{morgan_theory_1982}. As such MN reactions have been investigated in various systems (see e.g.~\cite{larson_mutual_2019} and references therein). 

Iodine plasma is one example of an electronegative plasma. Interest in iodine plasma has been renewed recently 
since it is a promising candidate to be used in electric propulsion systems, notably for satellites (see e.g.~\cite{rafalskyi_-orbit_2021,esteves_charged-particles_2022} and references therein). 

Very recently, the MN reaction between I$^+$ and I$^-$ ions has been studied experimentally~\cite{poline_final-state-resolved_2022}: the branching ratios for the different channels were measured at two collision energies, 0.1 eV and 0.8 eV. This work showed that the MN reaction forms iodine atoms either in their ground state or with one atom in an electronically excited state. These two classes of states were found to be populated with nearly equal proportions with no dependence on the collision energy. The total cross sections at these collision energies were estimated, but with fairly large uncertainties.

There is currently no accurate absolute cross sections published for the MN reaction between I$^+$ and I$^-$ ions, which impedes the modelling of iodine plasma. Investigating such collision system is a difficult task since iodine has a strong spin-orbit coupling and, moreover, the potential energy curves of I$_2$ exhibit multiple and overlapping avoided crossings, where the MN reaction can take place. The aim of the present work is to provide estimates of these cross sections in a broad range of collision energies. For that, we have employed a combination of \emph{ab initio} relativistic electronic structure calculations and the Laudau-Zener Surface Hopping (LZSH) method to compute the relevant cross sections. Our calculations are then compared to the recent experiments of \citeauthor{poline_final-state-resolved_2022} \cite{poline_final-state-resolved_2022}. 


This paper is organized as follows. In the next section we briefly outline the methods used in the present work. Sec. \ref{sec:Results} is devoted to the discussion of the theoretical results of this work and their comparison with the experimental results of \citeauthor{poline_final-state-resolved_2022} \cite{poline_final-state-resolved_2022}. The conclusions are reported in Sec. \ref{sec:conclu}. Atomic units are used throughout, unless explicitly indicated otherwise.

\section{Methods}
\label{sec:methods}

\subsection{Potential energy curves}
\label{sec:PEC}
\begin{figure*}[ht!]
        \centering
        \includegraphics[scale=0.50]{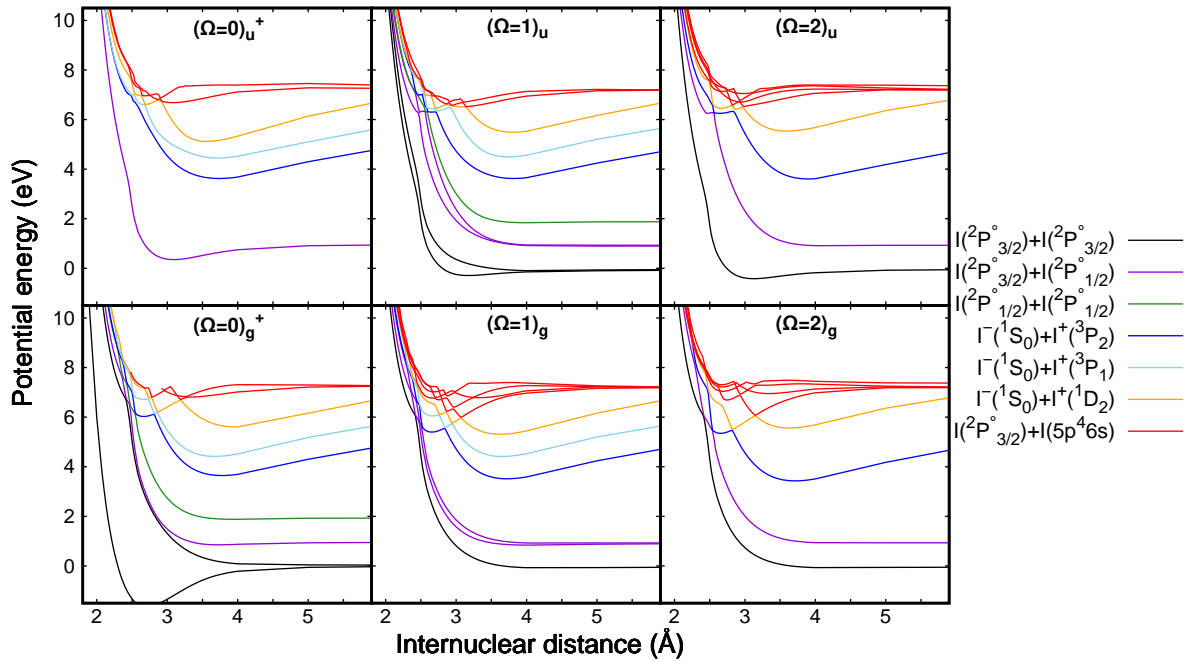}
        \caption{\label{fig:PEC} 
        Potential energy curves of 51 electronic states of I$_{2}$ computed with relativistic MRCI (reconstructed from the data from~\citeauthor{poline_final-state-resolved_2022} \cite{poline_final-state-resolved_2022}). Only the states of the symmetries that correlate with the $\text{I}^{-}(^{1}S_{0})+\text{I}^{+}(^{3}P_{2})$ reactants state are displayed
here.}
\end{figure*}

The potential energy curves used in this work, shown in figure~\ref{fig:PEC}, have been obtained with the multi-reference configuration interaction (MRCI) method, as implemented in the KRCI module~\cite{fleig_generalized_2003} of the DIRAC relativistic electronic structure package~\cite{saue_dirac_2020}. Such calculations have been carried out with the DIRAC19~\cite{gomes_dirac19_2019} release as well as with development version identified by hash \texttt{1e798e5}. We employed triple-zeta quality basis sets~\cite{dyall_relativistic_2006} supplemented by three diffuse functions so that Rydberg and ion-pair (IPr) states could be accurately represented. The reference wavefunction consisted of the set of determinants spanned by the $p^5$ manifold of each of the iodine atoms (thus representing 10 electrons in 12 spinors).
For further information, readers can consult the computational details section of~\citeauthor{poline_final-state-resolved_2022} \cite{poline_final-state-resolved_2022}. 

It should be noted that we computed the potential energy curves for states with 
projection of total electronic angular momentum $\Omega = 0, 1, 2$ but not for states with
$\Omega>2$ since the $\text{I}^{-}(^{1}S_{0})+\text{I}^{+}(^{3}P_{2})$ reactant state does not correlate with such states. Indeed the $\text{I}^{-}(^{1}S_{0})+\text{I}^{+}(^{3}P_{2})$ state correlates with states having the following angular momenta \cite{lukashov_iodine_2018}: the double degenerate $\left(\Omega=1\right)_{g}$, $\left(\Omega=1\right)_{u}$, $\left(\Omega=2\right)_{g}$ and $\left(\Omega=2\right)_{u}$, and the singly degenerate $\left(\Omega=0\right)_{g}^{+}$
and $\left(\Omega=0\right)_{u}^{+}$. 

Furthermore, for implementation reasons the KRCI module does not take into account the +/- symmetry and
thus is not able to differentiate directly the + and the - states.
In order to do that we also computed the dipole transition moments
between the $\left(\Omega=0\right)_{g}$ and the $\left(\Omega=0\right)_{u}$
states. Knowing that the lowest $\left(\Omega=0\right)_{g}$ state
is of + symmetry and that the lowest $\left(\Omega=0\right)_{u}$
state is of - symmetry \cite{lukashov_iodine_2018}, we were able to rebuild the
potential energy curves of the $\left(\Omega=0\right)_{g}^{+}$ and
$\left(\Omega=0\right)_{u}^{+}$ states using the selection rule stating
that the dipole transition between a + and a - state is
forbidden~\cite{banwell_fundamentals_1983}.

\subsection{Landau Zener Surface Hopping}
\label{sec:LZSH}

An accurate description of MN reactions at low collision energies requires, in principle, a fully quantum mechanical approach for the nuclear dynamics. However, in the current system, the potential energy curves of I$_2$ exhibit multiple and overlapping avoided crossings such that such a sophisticated approach is out of reach from a computational point of view.   To overcome this difficulty, in this work we employ the LZSH method \cite{belyaev_landau-zener_2014} to obtain the cross sections of the I$^{+}$+ I$^{-}$ mutual neutralization reaction. 

LZSH is a probabilistic, semi-classical method
in which the system is moving classically along the potential energy
curves. The non-adiabatic interactions are considered only at the
vicinity of avoided crossings \cite{desouter-lecomte_chapitre_2017,wigner_structure_2000}. The list of the avoided crossings considered in this work is given in appendix. Note that, as previously explained in \citeauthor{poline_final-state-resolved_2022} \cite{poline_final-state-resolved_2022}, we use a semi-empirical model \cite{olson_ionion_1970} to estimate the electronic coupling at large-distance ($R>7\mathring{A}$) crossings between the ion-pair states and the $\text{I}(5p^{4}6s)+\text{I}(^{2}P_{1/2})$ states. These couplings have been shown to be negligible. 

The LZSH method can be described as follows: The system starts at a distance $R_{0}$ on the curve corresponding to the reactants (\emph{i.e.} the curves which correlate with the $\text{I}^{-}(^{1}S_{0})+\text{I}^{+}(^{3}P_{2})$ ion pair state), $R_{0}$ being larger than
the internuclear distances of all avoided crossings (in this work, 
$R_{0}$ = 12 a.u.). The system then moves along this curve while it has
sufficient kinetic energy and until it reaches an avoided crossing. At
this point there is a probability $p_{\alpha\rightarrow\beta}^{LZ}$(given
by the Landau-Zener formula\cite{zener_non-adiabatic_1932,belyaev_nonadiabatic_2011}, equation \ref{eq:LZ}) that the system
hops from its starting state (named $\alpha$) to the other state
involved in the avoided crossing (named $\beta$), if its kinetic energy
is sufficient. We have :

\begin{equation}
p_{\alpha\rightarrow\beta}^{LZ}=exp\left(-\frac{\pi}{2v}\sqrt{\frac{\Delta V_{\alpha\beta}^{3}}{\frac{d^{2}}{dR^{2}}(\Delta V_{\alpha\beta})}}\right)\label{eq:LZ}
\end{equation}
where $v$ is the speed of the system at the crossing and $\Delta V_{\alpha\beta}$
is the energy difference between the two adiabatic potential
energy curves at the avoided crossing. $v$ is simply obtained by energy conservation:
\begin{equation}
v=\sqrt{\frac{2\left(E_{m}-V_{\alpha}(R)\right)}{\mu}}\label{eq:vitesse}
\end{equation}
with $\mu$ being the reduced mass of the system (for I$_{2}$, $\mu$ = 115666 a.u.).
$V_{\alpha}(R)$ is the adiabatic potential energy
curve $\alpha$ at the internuclear distance $R$, and $E_{m}$ the
mechanical energy of the system :
\begin{equation}
E_{m}=E_{coll}+V_{asymp}
\end{equation}
where $E_{coll}$ is the collision energy and $V_{asymp}$ is the energy of the $\text{I}^{-}(^{1}S_{0})+\text{I}^{+}(^{3}P_{2})$ reactant state at $R\rightarrow+\infty$.

When the kinetic energy of the system reaches 0, the system turns back, and when
it reaches $R_{0}$ again, the trajectory ends. By computing
a sufficiently high number of trajectories, we can compute a reaction
probability $P_{f}$ towards each of the possible product states
$f$. 
\begin{equation}
P_{f}=\frac{N_{f}}{N_{tot}}
\end{equation}
where $N_{f}$ is the number of trajectories which ended in the
product state $f$ and $N_{tot}$ is the total number of trajectories. In this work we used $N_{tot}=400$. We found that using a higher value of $N_{tot}$ has no significant impact on the results.

The cross sections towards each product state are then obtained by
integrating the $P_{f}$ over the angular momentum $l$ \cite{child_molecular_1974} :
\begin{equation}
\sigma_{f}^X(E_{coll})=\frac{\pi}{2\mu E_{coll}}\sum_{l=0}^{l=+\infty}(2l+1)P_{f}(E_{coll},l)\label{eq:SE}
\end{equation}
where $X$ denotes a given symmetry state of the I$_2$ potential energy curves.

To compute {$P_f(E_{coll},l)$ with $l\neq0$,
we are using the method described in the beginning of this paragraph
but replacing $V_{\alpha}(R)$ in the equation \ref{eq:vitesse} by
$V_{\alpha,eff}(R,l)$ to account for the rotational barrier :}

\begin{equation}
V_{eff,\alpha}(l,R)=V_{\alpha}(R)+\frac{l(l+1)}{2\mu R^{2}} \label{eq:V_eff}
\end{equation}

Practically, the sum in the equation \ref{eq:SE} stops (at a value
$l=l_{max}$) when the rotational barrier becomes too important for
the system to reach the farthest avoided crossing involving the reactant
state. We have :
\begin{equation}
l_{max}=-\frac{1}{2}+\sqrt{\frac{1}{4}-\frac{\mu R_{c}^{2}}{2}\left(4V(R_{c})-V_{asymp}-E_{coll}\right)}
\end{equation}
where $R_{c}$ and $V(R_{c})$ are the internuclear distance and the adiabatic energy of the reactant state at this avoided crossing.

This approach is used for each of the symmetries considered in this work
(see section \ref{sec:PEC}), the reaction cross sections towards each state
are then obtained by averaging over all symmetries, taking into account
their multiplicity, hence:
\begin{equation}
\sigma_{f}\left(E_{coll}\right)=\frac{\sum_{X\in symmetries}m_{X}\sigma_{f}^{X}\left(E_{coll}\right)}{\sum_{X\in symmetries}m_{X}\label{eq:sym}}
\end{equation}
with $m_{X}$ being the multiplicity of the symmetry $X$ and $\sigma_{\alpha}^{X}$
the reaction cross section towards the state $\alpha$ for the symmetry
$X$ obtained with equation \ref{eq:SE}.

\subsection{Empirical correction to the asymptotic energies}
\label{sec:Empirical corr}

When comparing the asymptotic energies of the reactant $\text{I}^{-}(^{1}S_{0})+\text{I}^{+}(^{3}P_{2})$
state and the asymptotic energies of the product $\text{I}(^{2}P_{3/2})+\text{}^{2}[2]$
state  (the lowest of the $\text{I}(^{2}P_{3/2})+\text{I}(5p^46s)$ states, see \cite{kramida_nist_1999}), both computed by the MRCI method, we noticed an inversion in
the energetic order with respect to the experimental energy levels
\cite{a_kramida_notitle_2019}. Thus, while the reaction path $\text{I}^{-}(^{1}S_{0})+\text{I}^{+}(^{3}P_{2})\rightarrow{}\text{I}(^{2}P_{3/2})+{}^{2}[2]$
should be open (even with a collision energy of 0) according to experimental data, from our calculations it is closed.

To correct for this qualitative and quantitative failure we decided to artificially add the kinetic energy
$\varepsilon$ to the system at the beginning of the trajectories
on the $\text{I}^{-}(^{1}S_{0})+\text{I}^{+}\left(^{3}P_{2}\right)$
reactant states so that the following equality is verified :
\begin{equation}
\begin{array}{l}
V_{exp}^{\infty}\left(\text{I}(^{2}P_{3/2})+\text{I}({}^{2}[2])\right)-V_{exp}^{\infty}\left(\text{I}^{-}(^{1}S_{0})+\text{I}^{+}(^{3}P_{2})\right) \\
\quad = V_{MRCI}^{\infty}\left(\text{I}(^{2}P_{3/2})+\text{I}({}^{2}[2])\right) \\
\qquad - \left[V_{MRCI}^{\infty}\left(\text{I}^{-}(^{1}S_{0})+\text{I}^{+}(^{3}P_{2})\right)+\varepsilon\right]
\end{array}
\end{equation}
where $V_{MRCI}^{\infty}(X)$ and $V_{exp}^{\infty}(X)$ denote respectively the asymptotic MRCI and experimental energies of the $X$ state. The value of $\epsilon$ is 0.63 eV. This is the only departure from the underlying \emph{ab initio} energy curves in our work.

\section{Results and discussion}
\label{sec:Results}

\begin{figure*}[ht!]
        \centering
        \includegraphics[scale=0.50]{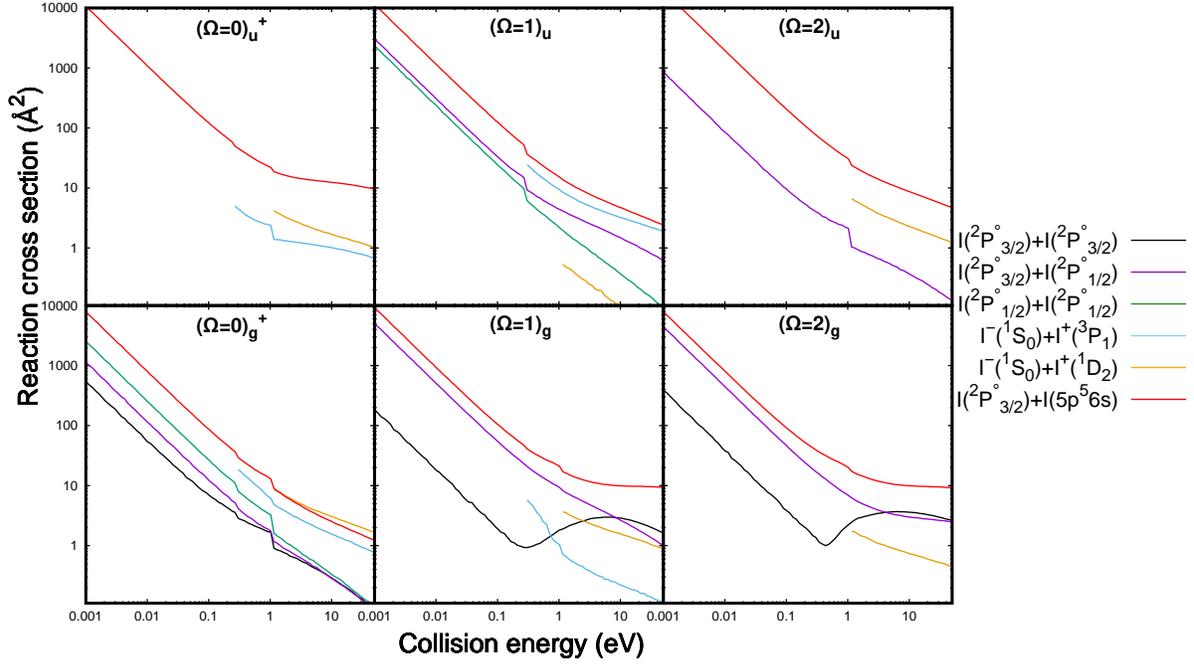}
        \caption{Cross sections for the reactions between $\text{I}^{+}$ and $\text{I}^{-}$ for the 6 symmetries correlating with the $\text{I}^{-}(^{1}S_{0})+\text{I}^{+}(^{3}P_{2})$ reactants state.\label{SE_sym}}
\end{figure*}

Using the potential energy curves presented in section \ref{sec:PEC},
we applied the LZSH method for each of the symmetries considered here
(see section \ref{sec:PEC}). We thus obtained the reaction cross sections towards
the following neutral product states : $\text{I}(^{2}P_{3/2})+{}\text{I}(^{2}P_{3/2})$, $\text{I}(^{2}P_{3/2})+{}\text{I}(^{2}P_{1/2})$, $\text{I}(^{2}P_{1/2})+{}\text{I}(^{2}P_{1/2})$ and $\text{I}(5p^{4}6s)+{}\text{I}(^{2}P_{1/2})$.

Here, we did not try to differentiate the different substates constituting
the $\text{I}(5p^{4}6s)$ configuration obtained with the MRCI method, since the energy difference between some of 
these substates is below 0.2 eV \cite{a_kramida_notitle_2019}. We lack extensive benchmark studies between MRCI and other approaches such as those based on coupled cluster wavefunctions for the iodine systems. However, from recent examples in the literature \cite{yuan_reassessing_2022, pototschnig_electronic_2021, denis_theoretical_2015} in which a comparison of methods has been made on an equal footing (same basis set and Hamiltonian), we see that among different correlated approaches, the corresponding electronic state energies can differ by values which are similar to, or higher than ,the differences among substates seen here.

\begin{figure}
        \includegraphics[scale=0.50]{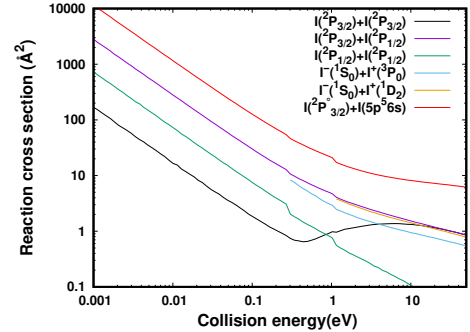}
        \caption{Total (symmetry averaged) cross sections for the $\text{I}^{+}+\text{I}^{-}\rightarrow2I$
        and $\text{I}^{+}+\text{I}^{-}\rightarrow\left(\text{I}^{+}\right)^{*}+\text{I}^{-}$ reactions. \label{SE_avg}}
\end{figure}

We also obtained the reaction cross
sections towards the two lowest energy excited ion-pair states $\text{I}^{-}(^{1}S_{0})+\text{I}^{+}\left(^{3}P_{1}\right)$
and $\text{I}^{-}(^{1}S_{0})+\text{I}^{+}\left(^{1}D_{2}\right)$.  The evolution of these reaction cross
sections with respect to the collision energy is shown in figure \ref{SE_sym} and the total symmetrized reaction cross sections, obtained with
equation \ref{eq:sym}, are shown in figure \ref{SE_avg}.

At collision energies lower than 0.1 eV the cross sections towards the neutral product states follow an asymptotic behavior proportional to the inverse of the collision energy. At these energies, for all symmetries, the most abundant product  is the neutral $\text{I}(5p^{4}6s)+{}\text{I}(^{2}P_{1/2})$ product, followed by the three lowest energy neutral products $\text{I}(^{2}P_{3/2})+{}\text{I}(^{2}P_{3/2})$, $\text{I}(^{2}P_{3/2})+{}\text{I}(^{2}P_{1/2})$, $\text{I}(^{2}P_{1/2})+{}\text{I}(^{2}P_{1/2})$ in this order. 

At collision energies higher than 0.3 eV the cross sections towards the $\text{I}(^{2}P_{3/2})+{}\text{I}(^{2}P_{3/2})$ state increase up to the collision energy of 10 eV while the cross sections towards the $\text{I}(^{2}P_{1/2})+{}\text{I}(^{2}P_{3/2})$ and $\text{I}(5p^{4}6s)+{}\text{I}(^{2}P_{1/2})$ states decrease at a slower rate than for the collision energies below 0.1 eV. At collision energies higher than 0.1 eV the cross sections towards the $\text{I}(^{2}P_{1/2})+{}\text{I}(^{2}P_{1/2})$ state continue to decrease as the inverse of the collision energy so it becomes negligible compared to the other cross sections.

The reaction cross sections towards the $\text{I}^{-}(^{1}S_{0})+\text{I}^{+}\left(^{3}P_{1}\right)$
and $\text{I}^{-}(^{1}S_{0})+\text{I}^{+}\left(^{1}D_{2}\right)$ ion pair states have energy thresholds of, respectively, 0.26 eV and 1.1 eV. The values of these cross sections after their threshold are of the same order of magnitude as the one of the $\text{I}(^{2}P_{1/2})+{}\text{I}(^{2}P_{3/2})$ state.

\begin{figure}
        \includegraphics[scale=0.50]{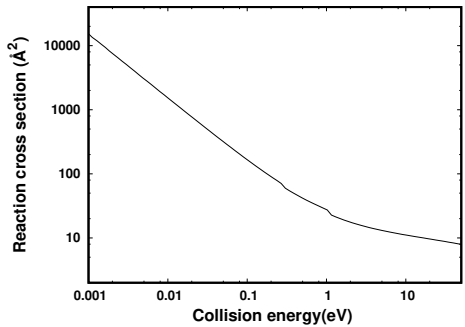}
        \caption{Total (symmetry averaged) cross section for mutual neutralization cross section between $\text{I}^{+}$ and $\text{I}^{-}$ \label{SE_neutr}}
\end{figure}

The total neutralization cross sections (sum of the cross sections toward all neutral states) is shown in figure \ref{SE_neutr}. It decreases as the inverse of the collision energy up to 0.1 eV and then decreases at a slower rate. The two discontinuities at 0.26 eV and 1.1 eV correspond to the energy thresholds of the reactions producing the two excited ion pairs.

In 2021, \citeauthor{poline_final-state-resolved_2022}\cite{poline_final-state-resolved_2022} conducted an experiment at the double ion storage ring DESIREE (Double ElectroStatic Ion Ring ExpEriment) in Stockholm. They were able to measure the branching ratios towards each of the neutral product states, more specifically, they obtained the ratio (denoted by $R_{\sigma}$) between the $\text{I}(5p^{4}6s)+{}\text{I}(^{2}P_{1/2})$ states and the $\text{I}(^{2}P_{3/2})+{}\text{I}(^{2}P_{3/2})$, $\text{I}(^{2}P_{3/2})+{}\text{I}(^{2}P_{1/2})$ and $\text{I}(^{2}P_{1/2})+{}\text{I}(^{2}P_{1/2})$ states, for collision energy of 0.1 and 0.8  eV. We therefore have $R_{\sigma}$ as :
{\small
\begin{equation}
R_{\sigma}=\frac{\begin{array}{c}
\sigma\left(I\left(^{2}P_{3/2}\right)+I\left(^{2}P_{3/2}\right)\right)+\sigma\left(I\left(^{2}P_{3/2}\right)+I\left(^{2}P_{1/2}\right)\right)\\
+\sigma\left(I\left(^{2}P_{1/2}\right)+I\left(^{2}P_{1/2}\right)\right)
\end{array}}{\begin{array}{c}
\sigma\left(I\left(5p^{4}6s\right)+I\left(^{2}P_{1/2}\right)\right)+\sigma\left(I\left(^{2}P_{3/2}\right)+I\left(^{2}P_{3/2}\right)\right)\\
+\sigma\left(I\left(^{2}P_{3/2}\right)+I\left(^{2}P_{1/2}\right)\right)+\sigma\left(I\left(^{2}P_{1/2}\right)+I\left(^{2}P_{1/2}\right)\right)
\end{array}}
\end{equation}

}

\begin{figure}
        \includegraphics[scale=0.50]{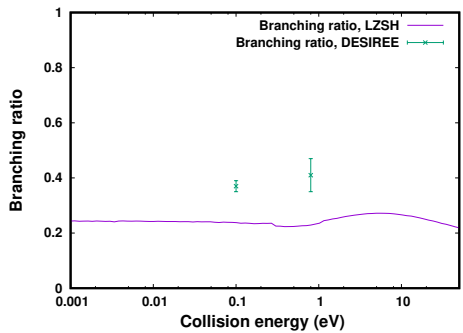}
        \caption{Comparison of the $R_{\sigma}$
ratios, computed with the LZSH method and measured by \citeauthor{poline_final-state-resolved_2022} \cite{poline_final-state-resolved_2022} \label{BR}}
\end{figure}

We can directly obtain this ratio from our calculations. The comparison between the theoretical ratio and the measurements is shown in figure \ref{BR}. Our results show that this branching ratio does not vary significantly with respect to the collision energy, with values between 22\% and 27\%. The measured and computed ratio are of the same order of magnitude. However, the LZSH-based model underestimates this ratio by a factor 1.5.

Moreover, our model gives a semi-quantitative agreement for the prediction of the ratios between the cross sections of the $\text{I}(^{2}P_{3/2})+{}\text{I}(^{2}P_{3/2})$, $\text{I}(^{2}P_{3/2})+{}\text{I}(^{2}P_{1/2})$ and $\text{I}(^{2}P_{1/2})+{}\text{I}(^{2}P_{1/2})$ states. These ratios, in comparison with those obtained by \citeauthor{poline_final-state-resolved_2022} \cite{poline_final-state-resolved_2022}, are displayed in table \ref{tab:ratios}.

\begin{table}
\begin{tabular}{ccccc}
\hline 
 & \multicolumn{2}{c}{$0.1$ eV} & \multicolumn{2}{c}{$0.8$ eV}\tabularnewline
\hline 
Product channel & LZSH & exp. & LZSH & exp.\tabularnewline
\hline 
$\text{I}(^{2}P_{3/2})+{}\text{I}(^{2}P_{3/2})$ & 5\% & 31\% & 12\% & 28\%\tabularnewline
$\text{I}(^{2}P_{3/2})+{}\text{I}(^{2}P_{1/2})$ & 76\% & 57\% & 76\% & 51\%\tabularnewline
$\text{I}(^{2}P_{1/2})+{}\text{I}(^{2}P_{1/2})$ & 19\% & 11\% & 13\% & 21\%\tabularnewline
\hline 
\end{tabular}\caption{Ratios of the cross sections between the three lowest neutral product states, obtained with the LZSH method and experimentally by \citeauthor{poline_final-state-resolved_2022} \cite{poline_final-state-resolved_2022} at collision energy of 0.1 and 0.8 eV.} 
\label{tab:ratios}
\end{table}

\citeauthor{poline_final-state-resolved_2022} \cite{poline_final-state-resolved_2022} were also able to estimate the absolute neutralization cross section, at a collision energy of 0.1 eV, to be in the range of $10^{3\pm1}$~\AA$^{2}$. Our results displayed in figure \ref{SE_neutr} (165~\AA$^{2}$ at 0.1 eV) agrees with this estimation. 

The disagreement between the experiments at DESIREE and our results may be attributed to the semiclassical approach employed in this work. However, given the complexity of the studied collisional system and the lack of data on the considered MN reaction such semi-quantitative estimates represent a significant step toward a better modelling, and thus understanding, of iodine plasma.

In order to gain more insights into the dynamics of the MN reaction, we investigate which avoided crossings are the ones which contribute
the most to the reactivity. We computed statistically the population
on each state as a function of time ($n_{\alpha}\left(t\right)$) :

\begin{equation}
n_{\alpha}\left(t\right)=\frac{N_{\alpha}\left(t\right)}{N_{traj}}
\label{eq:pop_time}
\end{equation}
where  $N_{\alpha}\left(t\right)$ is the number of trajectories being on the state $\alpha$ at the time $t$. Since the time does not appear explicitly in the method described in section \ref{sec:LZSH}, we computed it \emph{a posteriori}, by integrating Newton's law of motion (see equation \ref{eq:time_Newton},with $r_{i}$, $r_{j}$ being two adjacent points of the potential energy surface and $v$ being the speed of the system). The time is set arbitrarily at 0 when a trajectory starts at $R_{0}$. $N_{traj}$ is the total number of computed trajectories.
{\small
\begin{equation}
\begin{array}{c}
\Delta t_{ji}=\frac{\sqrt{2\mu}}{B}\left(\sqrt{E_{m}-V\left(r_{j}\right)}-\sqrt{E_{m}-V\left(r_{i}\right)}\right)~if~R~decreases\\
\Delta t_{ij}=\frac{\sqrt{2\mu}}{B}\left(\sqrt{E_{m}-V\left(r_{i}\right)}-\sqrt{E_{m}-V\left(r_{j}\right)}\right)~if~R~increases\\\
\end{array}
\label{eq:time_Newton}
\end{equation}
}
with $\Delta t_{lk} = t\left(r_{l}\right) - t\left(r_{k}\right)$ and 
$B=(V\left(r_{j}\right)-V\left(r_{i}\right))/(r_{j}-r_{i})$.

\begin{figure*}[t]
\centering
\hfill{}\includegraphics[width=8cm]{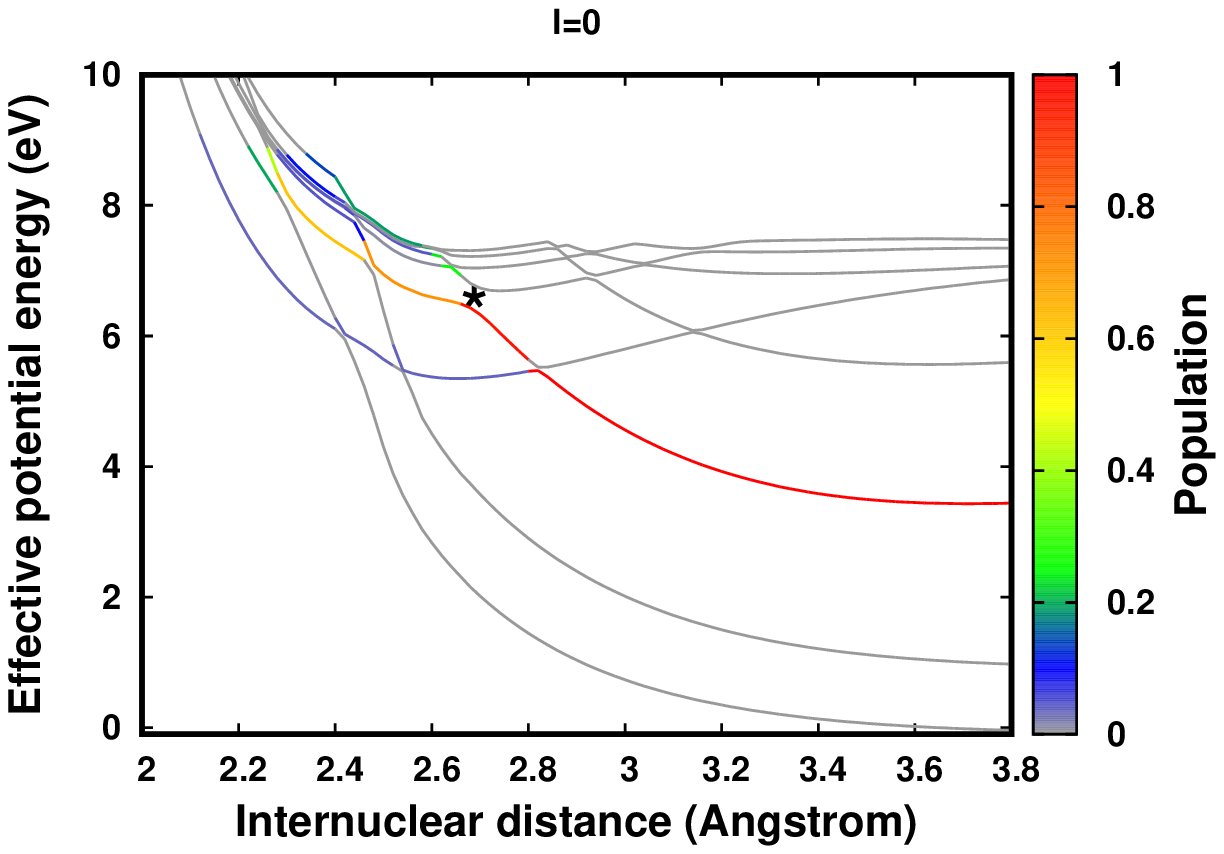}\hfill{}\includegraphics[width=8cm]{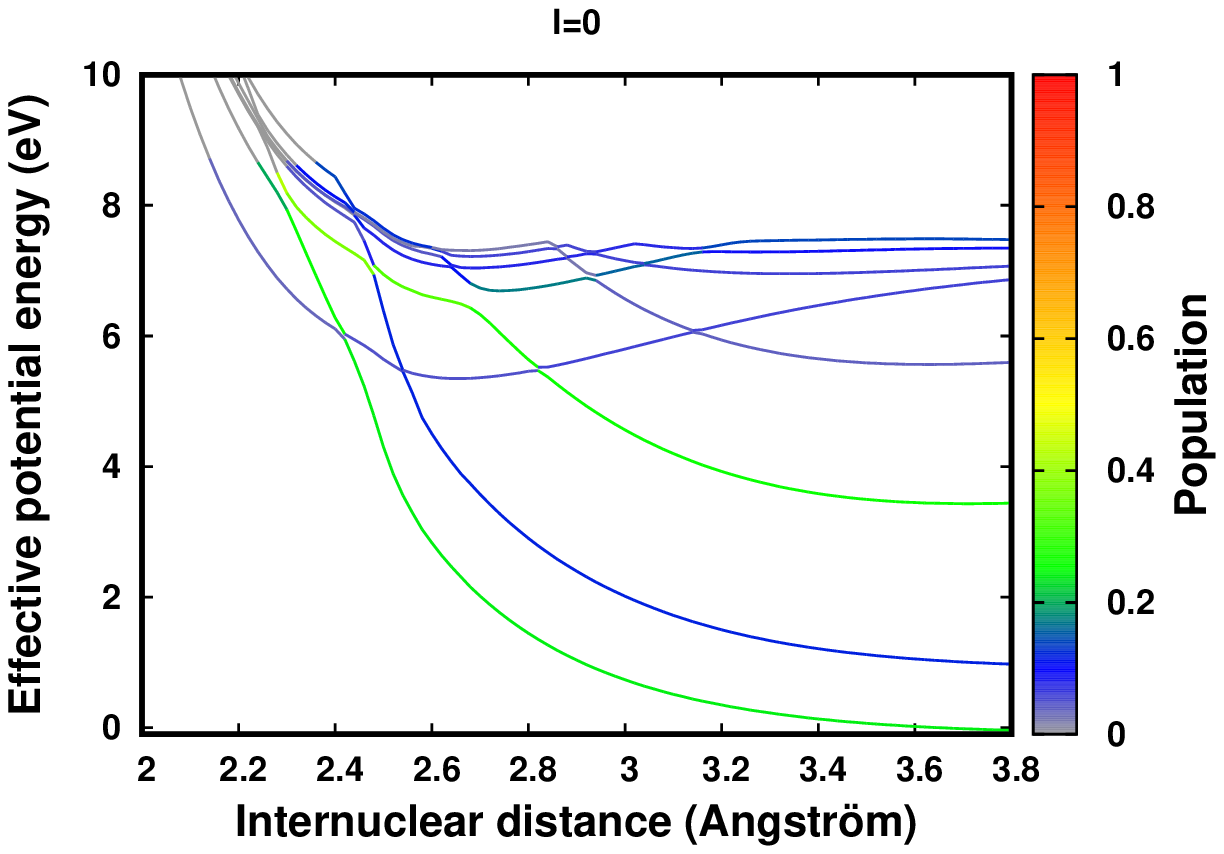}\hfill{}

\hfill{}\includegraphics[width=8cm]{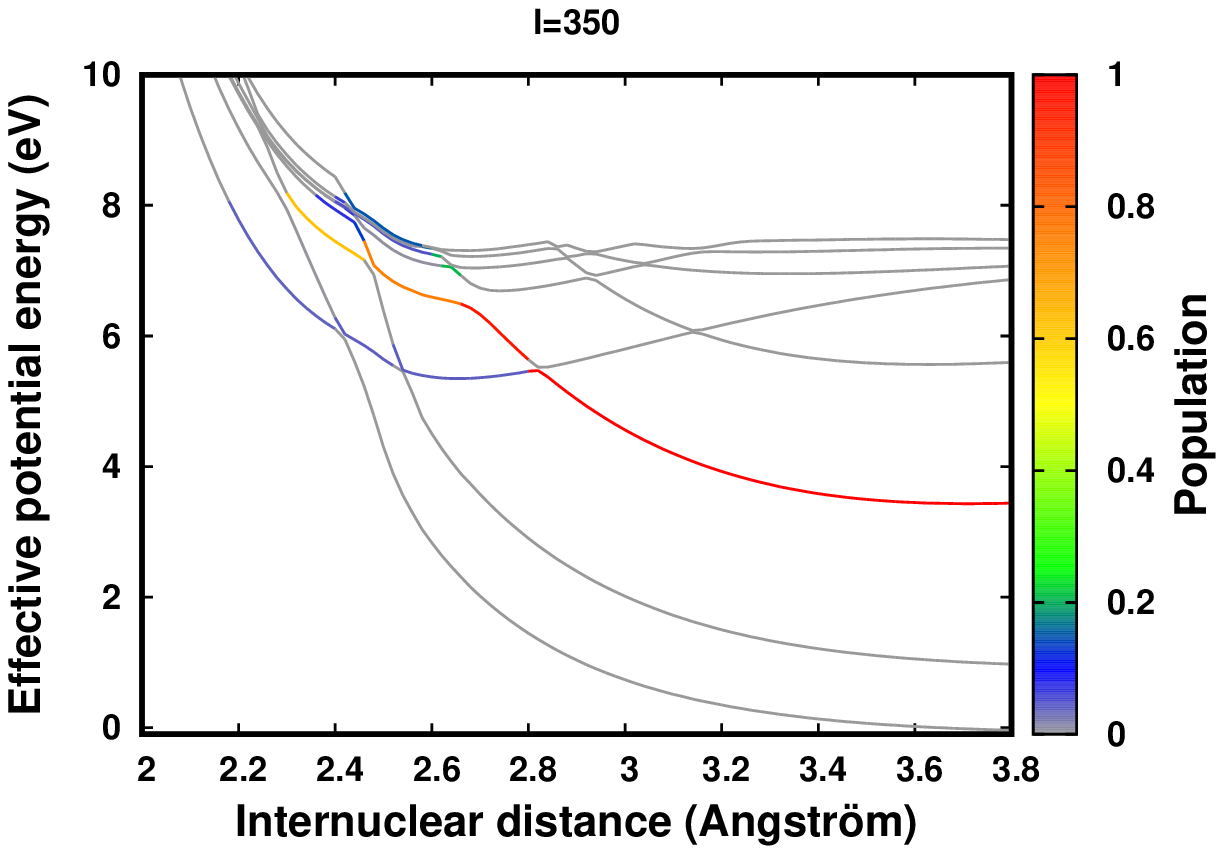}\hfill{}\includegraphics[width=8cm]{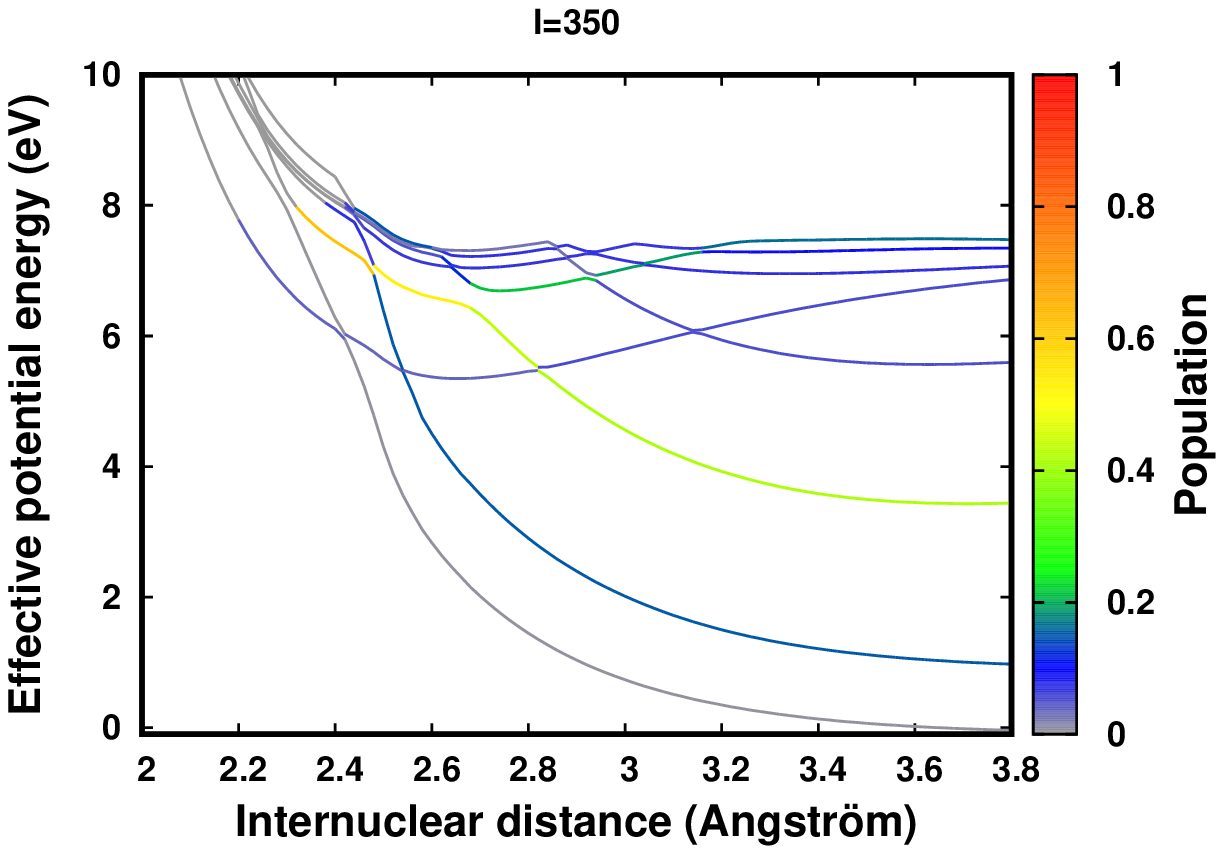}\hfill{}

\hfill{}\includegraphics[width=8cm]{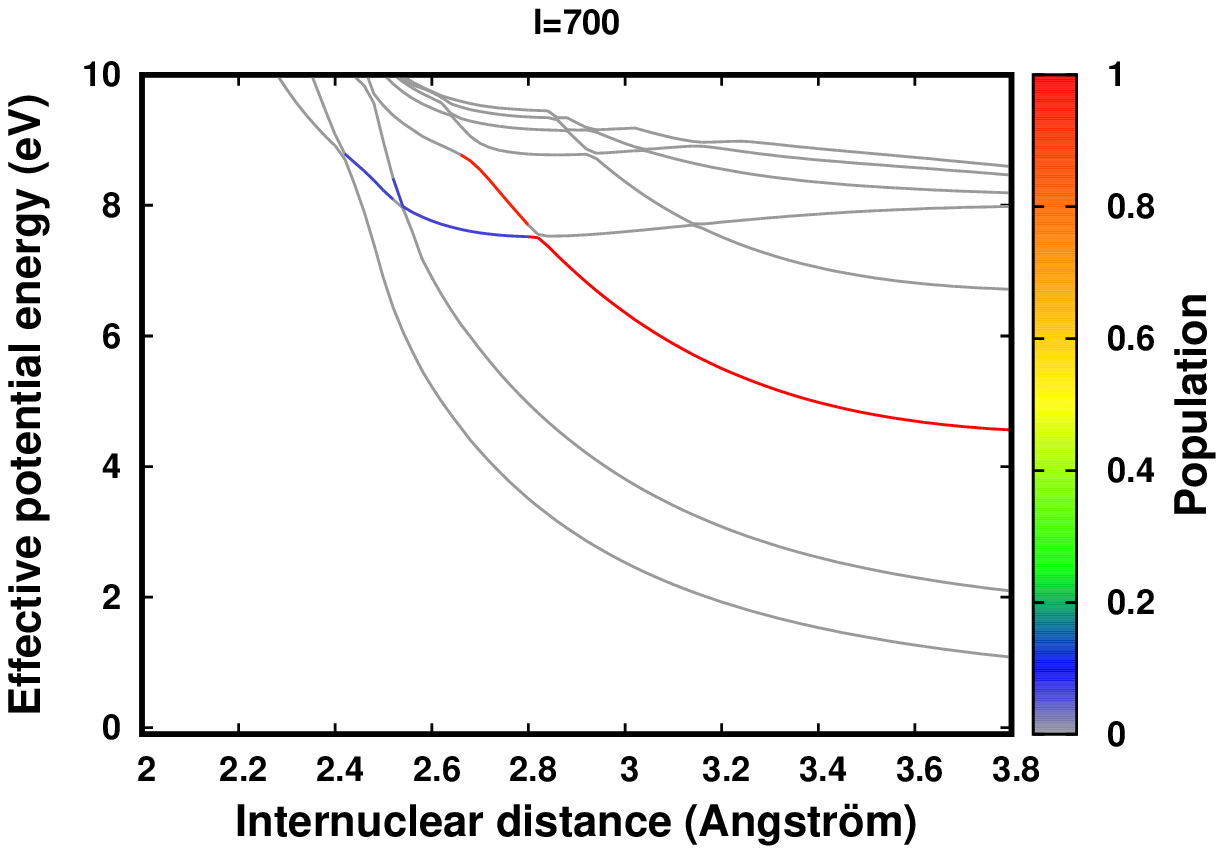}\hfill{}\includegraphics[width=8cm]{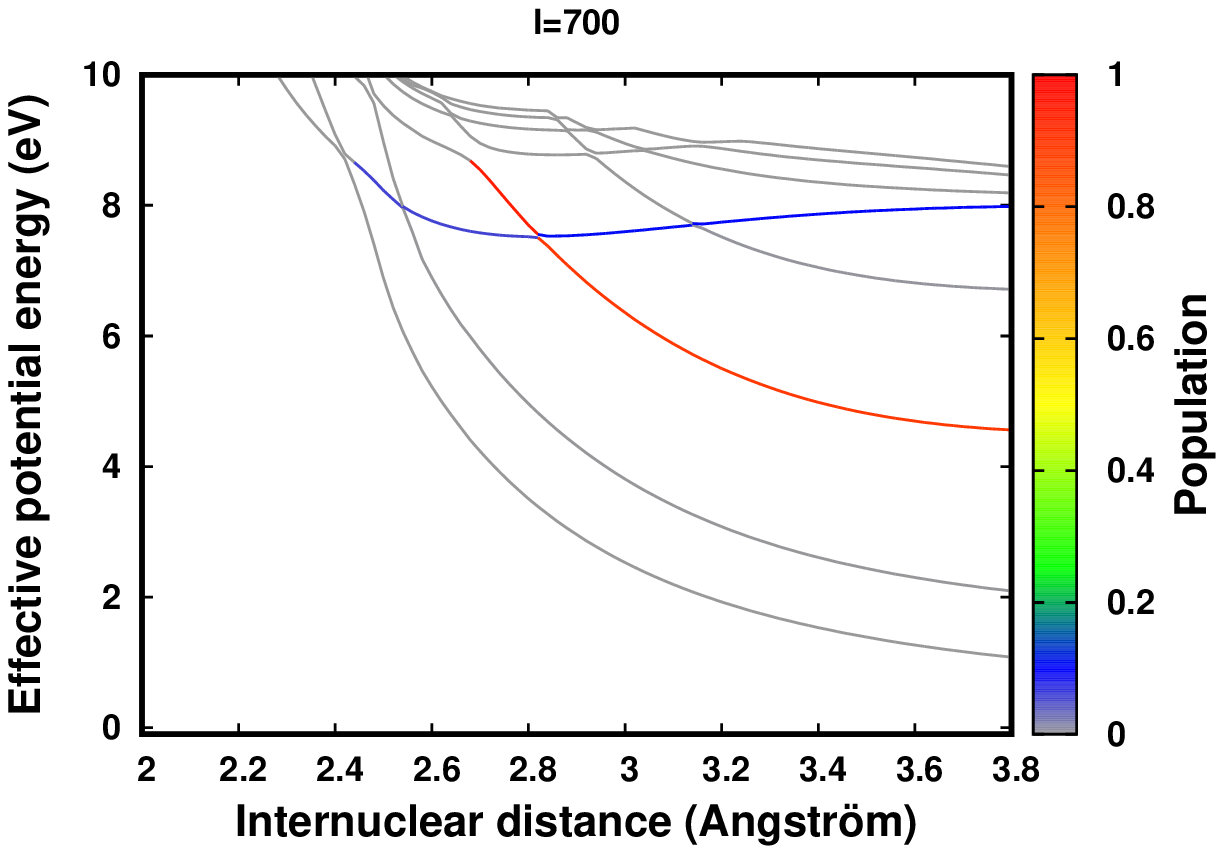}\hfill{}

\caption{Effective potential energy curves of the 8 first states of the $\left(\Omega=2\right)_{g}$ symmetry for 3 different values of angular momentum. The population $n_{\alpha}^{\leftarrow}\left(R\right)$ ($n_{\alpha}^{\rightarrow}\left(R\right)$) is displayed with a colorscheme in the left (right) panel. At $t=0$, the population is 1 in the lowest ion-pair state (the third state in energy order) and 0 in all the other states}
\label{fig:pop}
\end{figure*}

For each of the symmetries considered in this work we computed $N_{traj}=10000$
trajectories, for a collision energy of 0.9 eV and 3 different values
of the angular momentum $l$ ($l=0$, $l=350$ and $l=700$). The
populations obtained with these trajectories are then computed using
equation \ref{eq:pop_time} for each of the electronic states considered in this
work. The population on each of the 8 first electronic states of the
the $\left(\Omega=2\right)_{g}$ symmetry are displayed in figure
\ref{fig:pop}. For clarity we choose to represent separately the population
$n_{\alpha}^{\leftarrow}\left(R\right)$ coming from the part of the
trajectories with decreasing values of $R$ (before reaching the closest approach distance), and the population $n_{\alpha}^{\rightarrow}\left(R\right)$
coming from the part of the trajectories with increasing values of
$R$ (after reaching the closest approach distance), which are given by 
\begin{equation}
\begin{aligned}n_{\alpha}^{\leftarrow}\left(R\right)=\frac{N_{\alpha}^{\leftarrow}\left(R\right)}{N_{tot}} & \quad and\quad n_{\alpha}^{\rightarrow}\left(R\right)=\frac{N_{\alpha}^{\rightarrow}\left(R\right)}{N_{tot}}\end{aligned}
\label{eq:}
\end{equation}
 $N_{\alpha}^{\leftarrow}\left(R\right)$ ($N_{\alpha}^{\rightarrow}\left(R\right)$)
being the number of trajectories crossing the internuclear distance $R$
before (after) reaching the closest approach distance. In figure \ref{fig:pop}, 
$n_{\alpha}^{\leftarrow}\left(R\right)$ ( $n_{\alpha}^{\rightarrow}\left(R\right)$)
is shown in the left (right) panel of the figure using a color
scheme traced on the effective potential energy curves (see equation
\ref{eq:V_eff}) of the $\left(\Omega=2\right)_{g}$ symmetry.

At the first avoided crossing reached by the system (at $2.8\mathring{A}$),
it mainly has a diabatic behavior with approximately $90\%$ of the
population transferred to the higher energy state. This behavior is
observed for the majority of the avoided crossings of the system with
the important exception of the crossing between the fourth and
fifth states (in increasing energy order) at 2.7~\AA~(marked with a star in figure \ref{fig:pop}).
For this crossing we mainly observe an adiabatic behavior but still
with an important percentage of the population (about $30\%$) transferred
to the higher energy state. This intermediate behavior is directly
responsible of the reactivity towards the $\text{I}^{*}$ states, and indirectly
to the reactivity towards the lowest energy states through the avoided
crossings between the third and fourth states at 2.5~\AA~ and between the second and third states at 2.3~\AA.
The path towards the lowest energy states is the first to be screened
by the rotational barrier. A chemical reaction towards those states
is thus only possible for collisions with a low impact parameter (the
link between the impact parameter $b$ and the angular momentum $l$
is given by : $l=\sqrt{2\mu E_{coll}}*b$ \cite{child_molecular_1974}). 

The reactions
towards the $\text{I}(5p^{4}6s)+\text{I}(^{2}P_{3/2})$ states are still possible at higher values of $l$, which explains the higher reactivity towards those states (see figure \ref{BR}). The populations were also computed for the other symmetries. We did not find any major difference in the behavior of the populations between the $\left(\Omega=2\right)_{g}$ symmetry and the other symmetries.

\section{Conclusion}
\label{sec:conclu}

As a first step towards the generation of accurate models for the reactivity in iodine plasmas, in this work we have investigated a computational protocol, combining four-component multireference CI calculations for the I$_2$ system to obtain potential energy curves and the semiclassical Landau Zener surface hopping method to treat nuclear dynamics, to obtain theoretical cross sections of the mutual neutralization reaction between I$^{+}$ and I$^{-}$ for collision energies varying from 0.001 eV to 50 eV.  

Our results agree with the recent experimental measurements performed at the double ion ring DESIREE facility in the overlapping collision energy range. Furthermore, our work provides absolute cross sections over a broad range of collision energy. Our results show that the total cross section decrease from 1000\AA$^2$ at 0.001 eV collision energy to about 10 \AA$^2$ at 10 eV impact energy. Moreover, the branching ratios towards the different final states do not vary significantly with respect to the collision energy. We also studied the dynamics of this mutual neutralization reaction. 

The data and insights provided in this work will allow to model, beyond the current state of the art, the chemistry taking place in iodine plasma, which is particularly relevant for electric space propulsion.

\section{Acknowledgements}

SB thanks the École normale supérieure Paris-Saclay for the  PhD scholarship. XY and ASPG acknowledge funding from projects CompRIXS (ANR-19-CE29-0019, DFG JA 2329/6-1), Labex CaPPA (ANR-11-LABX-0005-01) and the I-SITE ULNE project OVERSEE and MESONM International Associated Laboratory (LAI) (ANR-16-IDEX-0004), and support from the French national supercomputing facilities (grants DARI A0090801859, A0110801859).
PLB ans NS thank Institut Physique des Infinis for financial support. SB, PLB and NS thank the members of the DESIREE facility for their warm welcome. The authors would also like to thank Anne Bourdon, Jean-Paul Booth and Benjamin Esteves for their interest in our work. 

\bibliography{biblio.bib}

\FloatBarrier

\clearpage

\section{Appendix}

\begin{table*}[ht!]
\scriptsize
\begin{tabular}{cccc}
Symmetry & Lower state & Higher state & R(u.a.)\tabularnewline
\hline 
\multirow{19}{*}{$(\Omega=0)_{g}^{+}$} & 3 & 4 & 4.35\tabularnewline
 & 2 & 3 & 4.46\tabularnewline
 & 5 & 6 & 4.46\tabularnewline
 & 4 & 5 & 4.51\tabularnewline
 & 1 & 2 & 4.57\tabularnewline
 & 3 & 4 & 4.57\tabularnewline
 & 2 & 3 & 4.65\tabularnewline
 & 5 & 6 & 4.69\tabularnewline
 & 4 & 5 & 4.74\tabularnewline
 & 3 & 4 & 4.80\tabularnewline
 & 2 & 3 & 5.02\tabularnewline
 & 5 & 6 & 5.18\tabularnewline
 & 6 & 7 & 5.25\tabularnewline
 & 4 & 5 & 5.32\tabularnewline
 & 5 & 6 & 5.41\tabularnewline
 & 7 & 8 & 5.59\tabularnewline
 & 7 & 8 & 5.96\tabularnewline
 & 6 & 7 & 6.01\tabularnewline
 & 7 & 8 & 6.20\tabularnewline
\hline 
\multirow{12}{*}{$(\Omega=0)_{u}^{+}$} & 4 & 5 & 4.57\tabularnewline
 & 3 & 4 & 4.69\tabularnewline
 & 1 & 2 & 4.68\tabularnewline
 & 2 & 3 & 4.80\tabularnewline
 & 1 & 2 & 4.88\tabularnewline
 & 4 & 5 & 4.96\tabularnewline
 & 3 & 4 & 4.99\tabularnewline
 & 2 & 3 & 5.06\tabularnewline
 & 4 & 5 & 5.22\tabularnewline
 & 3 & 4 & 5.40\tabularnewline
 & 4 & 5 & 5.48\tabularnewline
 & 3 & 4 & 5.52\tabularnewline
\hline 
\multirow{33}{*}{$(\Omega=1)_{g}$} & 5 & 6 & 4.23\tabularnewline
 & 4 & 5 & 4.27\tabularnewline
 & 6 & 7 & 4.32\tabularnewline
 & 3 & 4 & 4.37\tabularnewline
 & 2 & 3 & 4.40\tabularnewline
 & 5 & 6 & 4.42\tabularnewline
 & 1 & 2 & 4.45\tabularnewline
 & 4 & 5 & 4.47\tabularnewline
 & 3 & 4 & 4.50\tabularnewline
 & 0 & 1 & 4.57\tabularnewline
 & 6 & 7 & 4.57\tabularnewline
 & 5 & 6 & 4.61\tabularnewline
 & 2 & 3 & 4.61\tabularnewline
 & 4 & 5 & 4.65\tabularnewline
 & 1 & 2 & 4.69\tabularnewline
 & 3 & 4 & 4.69\tabularnewline
 & 2 & 3 & 4.78\tabularnewline
 & 7 & 8 & 5.04\tabularnewline
 & 6 & 7 & 5.10\tabularnewline
 & 5 & 6 & 5.12\tabularnewline
 & 4 & 5 & 5.22\tabularnewline
 & 8 & 9 & 5.23\tabularnewline
 & 7 & 8 & 5.26\tabularnewline
 & 6 & 7 & 5.29\tabularnewline
 & 3 & 4 & 5.37\tabularnewline
 & 8 & 9 & 5.38\tabularnewline
 & 5 & 6 & 5.40\tabularnewline
 & 7 & 8 & 5.43\tabularnewline
 & 4 & 5 & 5.55\tabularnewline
 & 6 & 7 & 5.60\tabularnewline
 & 8 & 9 & 5.65\tabularnewline
 & 5 & 6 & 5.78\tabularnewline
 & 7 & 8 & 6.15\tabularnewline
\end{tabular}\hfill{}%
\begin{tabular}{cccc}
Symmetry & Lower state & Higher state & R(u.a.)\tabularnewline
\hline 
\multirow{23}{*}{$(\Omega=1)_{u}$} & 6 & 7 & 4.42\tabularnewline
 & 5 & 6 & 4.46\tabularnewline
 & 3 & 5 & 4.57\tabularnewline
 & 2 & 3 & 4.65\tabularnewline
 & 6 & 8 & 4.64\tabularnewline
 & 6 & 7 & 4.72\tabularnewline
 & 5 & 6 & 4.76\tabularnewline
 & 4 & 5 & 4.80\tabularnewline
 & 7 & 9 & 4.80\tabularnewline
 & 6 & 7 & 4.84\tabularnewline
 & 3 & 4 & 4.84\tabularnewline
 & 5 & 6 & 4.87\tabularnewline
 & 4 & 5 & 4.91\tabularnewline
 & 9 & 10 & 4.99\tabularnewline
 & 7 & 8 & 4.99\tabularnewline
 & 6 & 7 & 5.10\tabularnewline
 & 8 & 9 & 5.10\tabularnewline
 & 7 & 8 & 5.40\tabularnewline
 & 8 & 9 & 5.48\tabularnewline
 & 6 & 8 & 5.51\tabularnewline
 & 9 & 10 & 5.78\tabularnewline
 & 8 & 9 & 5.86\tabularnewline
 & 7 & 8 & 5.93\tabularnewline
\hline 
\multirow{21}{*}{$(\Omega=2)_{g}$} & 2 & 6 & 4.23\tabularnewline
 & 3 & 4 & 4.23\tabularnewline
 & 1 & 2 & 4.35\tabularnewline
 & 6 & 7 & 4.58\tabularnewline
 & 0 & 1 & 4.58\tabularnewline
 & 4 & 6 & 4.61\tabularnewline
 & 3 & 4 & 4.65\tabularnewline
 & 1 & 2 & 4.80\tabularnewline
 & 6 & 7 & 4.93\tabularnewline
 & 2 & 3 & 4.70\tabularnewline
 & 5 & 6 & 4.96\tabularnewline
 & 4 & 5 & 4.99\tabularnewline
 & 3 & 4 & 5.06\tabularnewline
 & 2 & 3 & 5.32\tabularnewline
 & 6 & 7 & 5.40\tabularnewline
 & 5 & 6 & 5.44\tabularnewline
 & 4 & 5 & 5.56\tabularnewline
 & 6 & 7 & 5.56\tabularnewline
 & 5 & 6 & 5.74\tabularnewline
 & 3 & 4 & 5.93\tabularnewline
 & 6 & 7 & 5.97\tabularnewline
\hline 
\multirow{19}{*}{$(\Omega=2)_{u}$} & 3 & 4 & 4.68\tabularnewline
 & 6 & 7 & 4.72\tabularnewline
 & 5 & 6 & 4.73\tabularnewline
 & 4 & 5 & 4.74\tabularnewline
 & 3 & 4 & 4.77\tabularnewline
 & 2 & 3 & 4.81\tabularnewline
 & 1 & 2 & 4.83\tabularnewline
 & 4 & 5 & 4.98\tabularnewline
 & 6 & 7 & 5.10\tabularnewline
 & 5 & 7 & 5.18\tabularnewline
 & 4 & 5 & 5.25\tabularnewline
 & 3 & 4 & 5.33\tabularnewline
 & 2 & 3 & 5.37\tabularnewline
 & 6 & 7 & 5.40\tabularnewline
 & 4 & 5 & 5.48\tabularnewline
 & 5 & 6 & 5.51\tabularnewline
 & 4 & 5 & 5.56\tabularnewline
 & 3 & 4 & 5.63\tabularnewline
 & 6 & 7 & 5.96\tabularnewline
\end{tabular}

\caption{List of the avoided crossings considered in this work. For each avoided crossing we give its internuclear distance and the index of the 2 electronic states concerned by this crossing. The electronic states indexes are given by their energetic order (starting from zero for the lowest energy state of each symmetry)}

\end{table*}

\FloatBarrier

\end{document}